# Anisotropic Magnetocaloric Properties of The Ludwigite Single Crystal $Cu_2MnBO_5$

A.G. Gamzatov[1,*], Y.S. Koshkid'ko[2], D. C. Freitas[3], E. Moshkina[4], L. Bezmaternykh[4]

A.M. Aliev[1], S.-C. Yu[5], and M.H. Phan[6]

[1]*Amirkhanov Institute of Physics, DSC of RAS, 367003, Makhachkala, Russia*

[2]*Institute of Low Temperature and Structure Research, PAS, 50-950, Wroclaw, Poland*

[3]*Instituto de Física, Universidade Federal Fluminense, Campus da Praia Vermelha, 24210-346 Niterói, RJ, Brazil*

[4]*L.V. Kirensky Institute of Physics SB RAS, 660036 Krasnoyarsk, Russia*

[5]*Department of Physics, Ulsan National Institute of Science and Technology, Ulsan 44919, South Korea*

[6]*Department of Physics, University of South Florida, 4202 East Fowler Avenue Tampa, Florida 33620, USA*

**Abstract**

We present the results of a thorough study of the specific heat and magnetocaloric properties of a ludwigite crystal $Cu_2MnBO_5$ over a temperature range of 60 – 350 K and in magnetic fields up to 18 kOe. It is found that at temperatures below the Curie temperature ($T_C \sim 92$ K), $C_P(T)/T$ possesses a linear temperature-dependent behavior, which is associated with the predominance of two-dimensional antiferromagnetic interactions of magnons. The temperature independence of $C_P/T=f(T)$ is observed in the temperature range of 95 – 160 K, which can be attributed to the excitation of the Wigner glass phase. The magnetocaloric effect (i.e. the adiabatic temperature change, $\Delta T_{ad}$ $(T,H)$) was assessed through a direct measurement or an indirect method using the $C_P(T,H)$ data. Owing to its strong magnetocrystalline anisotropy, an anisotropic MCE or the rotating MCE ($\Delta T_{ad}^{rot}$ $(T)$) is observed in $Cu_2MnBO_5$. A deep minimum in the $\Delta T_{ad}^{rot}$ $(T)$ near the $T_C$ is observed and may be associated with the anisotropy of the paramagnetic susceptibility.

*Corresponding authors: gamzatov_adler@mail.ru (A.G)

Ludwigites belong to quasi-low-dimensional transition metal oxyborates, which are prominent representatives of systems with strongly correlated properties [1-3]. The oxyborate $Cu_2MnBO_5$ has a monoclinic-distorted structure of ludwigite and is characterized by P21/c space group due to the presence of $Cu^{2+}$ and $Mn^{3+}$ cations in Jahn-Teller cations. Macroscopic magnetic and specific heat studies have shown that this compound undergoes a magnetic phase transition into the ferrimagnetic phase at $T_C \sim 92$ K [1-3]. The crystal-orientational dependences of magnetization and magnetic susceptibility revealed the anisotropic magnetic characteristic both below the $T_C$ and in the paramagnetic regime (associated with the anisotropy of g-tensor due to monoclinic distortions introduced by Jahn-Teller $Cu^{2+}$ and $Mn^{3+}$ ions [3]). The microscopic magnetic structure of this ludwigite was experimentally studied using powder neutron diffraction [2]. It was determined that out of four nonequivalent positions occupied by magnetic ions, three positions are predominantly occupied by $Cu^{2+}$ ions, and one by $Mn^{3+}$ ions. It was also found that in the compound under study there was only a partial ordering of the magnetic moments in the ferrimagnetic phase due to the small moment of $Cu^{2+}$ ion in position *2a*. The magnetic moments of $Cu^{2+}$ and $Mn^{3+}$ ions in the ferrimagnetic phase are antiparallel and their directions do not coincide with the main crystallographic directions in the crystal [2]. The structural, magnetic and thermodynamic properties of $Cu_2MnBO_5$ were studied [1-3]. However, the magneto-thermal response of the material has not been investigated to date.

The utilizing of the magnetocaloric effect (MCE) in magnetic cooling technology is of current interest as it has the potential to replace conventional gas compression techniques [4, 5]. In addition to its perspective cooling application, the MCE has recently been used as a useful research tool for the analysis and interpretation of competing magnetic phases and the collective magnetic phenomena in a wide range of exotic magnetic materials [6-10]. In this regard, we present here results of the first comprehensive study of the anisotropic and frequency-dependent magnetocaloric properties of the ludwigite crystal $Cu_2MnBO_5$.

The Cu$_2$MnBO$_5$ single crystals were synthesized by the flux method with the ratio of the initial components Bi$_2$Mo$_3$O$_{12}$:1.3B$_2$O$_3$:0.7Na$_2$CO$_3$:0.7Mn$_2$O$_3$:2.1CuO by spontaneous nucleation [3]. The Cu$_2$MnBO$_5$ sample has a plate shape, with dimension of 3x3x0.3 mm$^3$ in the *a\*b\*c* configuration. The specific heat was measured by an ac-calorimetry. Direct measurements of the adiabatic temperature change $\Delta T_{ad}$ were carried out by the modulation method [11] in the direction of the magnetic field in the *ab*-plane. The application of an alternating magnetic field to the sample induced a periodic change in the temperature of the sample, due to the MCE. This temperature change was recorded by a differential thermocouple. The frequency of the alternating magnetic field in this experiment was 0.2 Hz. An alternating magnetic field with amplitude of up to 4 kOe was generated using an electromagnet and an external control power unit. The control alternating voltage was supplied to the power unit from the output of the (Lock-in) amplifier SR 830. An alternating magnetic field of 18 kOe was created by a source of permanent magnetic field of an adjustable intensity manufactured by AMT & CLLC.

Fig. 1 shows the temperature dependence of the specific heat $C_p(T)$ for Cu$_2$MnBO$_5$. The inset of Fig.1 demonstrates the temperature dependence of $C_P/T$ in the range of 70–150 K and at $H = 0$, 6.2, and 11 kOe. The $C_p(T)$ dependence of Cu$_2$MnBO$_5$ was reported in [2], and our results are in good agreement with these data. Note that $C_P(T)$ dependence exhibits a pronounced lambda anomaly at $T_C = 92$ K, associated with the ferromagnetic-paramagnetic (FM-PM) phase transition, which is suppressed upon the application of a magnetic field, while the maximum of the heat capacity shifts by 4 K toward a higher temperature at $H = 11$ kOe. Two features of the $C_p(T,H)$ behavior are worthy of note. Below the $T_C$, the $C_P/T(T)$ is linear and unchanged with temperature in the temperature range of 95-160 K. A similar behavior was observed for a number of ferroborates [12, 13]. At $T < T_C$, the $C_P/T = f(T)$ is well described by the expression $C_P = \alpha T^2$ (the dashed line in Fig. 1). This behavior has been explained by the predominance of two-dimensional antiferromagnetic interactions of magnons [12].

Using the expressions, $C_P = \alpha T^2$, $\alpha = 7.2k_B^3/2\pi D(N/A)$, where $\alpha$ is determined from the experimental data (Fig. 1) $\alpha = 0.01118$ J/mol K$^3$, (N/A) = 3·10$^{-9}$ mole/cm$^2$ – the number of magnetic ions per unit area [12], we can estimate the spin wave stiffness $D = 7.2k_B^3/2\pi\alpha(N/A) = 8.9810^{-59} J^2 cm^2$. Then using $\sqrt{D} = zJSa$, we have obtained the numerical value of the exchange integral from the $C_P(T)$ data at $T<T_C$ as $J = \sqrt{D}/zSa = 5.2610^{-23} J$ or J/k$_B$ = 3.8 K. Here we suppose $S = 2$ for Mn$^{3+}$ ions, $z = 3$ and $a = 3$ Å as the medium distance of local ions. We have neglected little variations in distance between the local ions and the magnetic arrangements of the walls. This value of J/k$_B$ = 4 K, which is compared to that reported for Fe$_3$BO$_5$ (J/k$_B$ = 2.1 K), which orders near 70 K [12]. We can also obtain the magnons speed in the walls of Cu$_2$MnBO$_5$, $v = 8.9910^2\ m/s$, by considering a linear dispersion relation with k $\hbar w = ck$. Just to compare, 2D antiferromagnetic magnons were also found in another oxyborate called hulsite with $v = 1.1810^3\ m/s$ [14]. This reinforces the presence of 2D antiferromagnetic magnons in our compound.

The temperature-independent $C_P/T=f(T)$ behavior is observed in the temperature range of 95-160 K (Fig. 1), which is associated with precursor effects of the charge ordering in a Wigner glass phase [12]. Another possibility is that it may arise from short range magnetic ordering in a low dimensional structure persisting just above the magnetic transition [13]. It is interesting to point out that the coefficients of these linear terms, $C/T$ = 773 mJ/mol K$^2$ obtained for Fe$_3$BO$_5$ [12] and $C/T$ = 843 mJ/mol K$^2$ for the present work are of similar magnitude.

The inset of Fig. 1 also shows the temperature dependence of the magnetic contribution to the specific heat, $\Delta C_P(T)=C_P - C_{Ph}$ ($C_{Ph}$ is the background contribution/lattice contribution). The value of the heat capacity jump $\Delta C_p$ in the phase transition region is ≈8 J/mol K. The temperature dependence of $\Delta S$ is also shown in Fig. 1, associated with the disordering of the magnetic system during the phase transition and determined using the expression: $\Delta S(T) = \int(\Delta C_P/T)dT$. According to our data, the $\Delta S$ is 0.44 J/mol K, which is slightly smaller than that obtained by the authors of work [2] ($\Delta S \sim 0.6$ J/mol K). Both values are much smaller compared with the theoretical value $\Delta S^* = \Delta S_{Mn}+\Delta S_{Cu} = n_{Mn}R\ln(2S(Mn^{3+})+1)+n_{Cu}R\ln(2S(Cu^{2+})+1) = 25.2$ J/mol K [2]. Such difference in the experimental and theoretical estimates of $\Delta S^*$ is characteristic of many magnetic materials, due to various reasons such as magnetic inhomogeneity [15-17] or this difference is indicative of the absence of complete ordering of the magnetic moments at the magnetic phase transition, which agrees with the results from the neutron magnetic scattering data [2].

Direct measurements of MCE ($\Delta T_{ad}$) were carried out in magnetic fields up to 18 kOe. To our best knowledge, no information on direct MCE measurements was reported in ludwigites. In work [18], for $Co_{4.76}Al_{1.24}(O_2BO_3)_2$, the maximum value of the MCE was estimated from the $C_P(T,H)$ data. According to the data [18], at $T = 80$ K (far from the phase transition temperature $T_N = 57$ K), the maximum $\Delta S$ of 7.5 J/kg K was observed for a field change of 9 T.

Figure 2(a) compares values of $\Delta T_{ad}$ in a magnetic field of 18 kOe for two orientations of the crystal (see the inset of Fig. 3 (a)). The maximum value of $\Delta T_{ad}$, according to direct measurements in a magnetic field of 18 kOe, is $\Delta T_{\parallel} = 0.45$ K at $T \approx 95$ K. A rotation of 90 degrees leads to a decrease to $\Delta T_{\perp} = 0.30$ K, i.e. the magnitude of $\Delta T_{ad}$ decreases by 1.5 times. In addition, the maximum of $\Delta T_{ad}$ also shifts toward a low temperature $T = 93$ K. As can be seen from Fig. 3(b), the strong anisotropy of the MCE appears below the $T_C$. So, at $T = 82$ K, the ratio $\Delta T_{\parallel}/\Delta T_{\perp} = 6.6$, while at $T = T_C$ the ratio $\Delta T_{\parallel}/\Delta T_{\perp} = 1.5$. The observed anisotropy of the MCE is in good agreement with the anisotropy of the magnetization of $Cu_2MnBO_5$ [2].

Figure 2(b) shows the change in temperature $\Delta T_{rot}$ of $Cu_2MnBO_5$ due to a 90° rotation of the crystal within the *ab* plane, which is defined as the rotating MCE (RMCE). These values of $\Delta T_{rot}$ are obtained by subtracting the curves shown in Fig. 2(a), measured along the two directions of the magnetic field at $H\|a$ and at $H\perp a$, i.e. $\Delta T_{rot} = \Delta T_{\|a} - \Delta T_{\perp a}$. As one can see from Fig. 2(b), the $\Delta T_{rot}(T)$ dependence shows a minimum near the $T_C$, and the $\Delta T_{rot}$ reaches a maximum at $T = 82$ K, which is 0.24 K at $H = 18$ kOe. Although the $\Delta T_{rot}$ value is not gigantic, it is comparable with the $\Delta T_{H\perp a}$ value for this sample, as well as with that of the RMCE for Gd ($\Delta T_{rot} \sim 0.3$ K at 15 kOe) [19, 20]. The inset of Fig. 3b shows the angular dependence of $\Delta T_{rot}$ at $T = 80$ K in a field of 18 kOe. At present, the record values of $\Delta T_{rot}$ are -4.4 K at H = 20 kOe for DyNiSi [21], 5.2 K at $H = 50$ kOe for HoMn$_2$O$_5$ [22], and $-$ 1.6 K at $H = 13$ kOe for NdCo$_5$ [23]. It is also important to note that the RMCE is reversible (see the inset of Fig. 3 (b)), which is a necessary condition for a magnetocaloric material to be used in magnetic cooling technology. Balli *et al.* proposed [22, 24] a rotating magnetic refrigerator in which the regenerator consists of single-crystal blocks of HoMn$_2$O$_5$ or TbMn$_2$O$_5$, and is rotated in a constant magnetic field. The idea of creating a thermomagnetic generator based on the RMCE was proposed at an earlier time though [25].

The observed anomaly in $\Delta T_{rot}(T)$ near the $T_C$ on is also worthy of note. Such anomalous behavior was not observed previously when studying the anisotropy of MCE in magnetic single crystals. According to [26], an equation for determining the adiabatic change in temperature of a ferrimagnet taking into account the contribution of two magnetic sublattices can be written in the following form:

$$\Delta T_{ad} = \frac{T}{C_{P,H}} \left( \frac{\partial \vec{M_1}}{\partial T} d\vec{H} + \frac{\partial \vec{M_2}}{\partial T} d\vec{H} \right),$$

where $M_1$ is the magnetization of the first sublattice, $M_2$ is the magnetization of the second antiparallel sublattice, d$H$ is the magnetic field increment, and $C_{P,H}$ is the heat capacity. For Cu$_2$MnBO$_5$, these are the Mn$^{3+}$ and Cu$^{2+}$ sublattices, which determine positive and negative exchange interactions in the system. It is suggested that the minimum in $\Delta T_{rot}(T)$ near the $T_C$ is a combination of several mechanisms. Firstly, magnetostriction can also make a significant contribution to the MCE along with the paraprocess [26]. It was shown in [5] that the magnetostriction anisotropy of Cu$_2$MnBO$_5$ has an unusual form; for a parallel configuration of $H\|c$ the crystal in the fields under consideration up to 20 kOe is compressed along the $c$-axis. At the same time, for the $H\perp c$ configuration, the magnetostriction has a usual quadratic form (the crystal expands). It is known that magnetostrictive contributions to MCE, depending on whether the crystal is compressed or expanded, will have opposite signs [27]. Another mechanism causing the anomaly in $\Delta T_{rot}(T)$ near the $T_C$ could be related to the temperature shift of the MCE maxima at different orientations of the crystal in a magnetic field (see Fig. 3a), due to the competition of positive and negative exchange interactions in $Cu_2MnBO_5$.

As is known, the energy of magnetocrystalline anisotropy (MCA) is determined through the MCA constants, which determine the RMCE. According to [25], the magnetic contribution to entropy caused by the rotation of the magnetization vector is equal to:

$$\Delta S_{anis} = -\frac{\partial \Delta E_{anis}}{\partial T} \qquad (1)$$

where $\Delta E_{anis}$ - change in the anisotropy energy upon rotation of the sample in a magnetic field from the *b*-axis to the *a*-axis. For the case of rotation of the magnetization vector in the *ab*-plane near Curie temperature, taking into account the first anisotropy constant, the change in the anisotropy energy can be written in the following form:

$$\Delta E_{anis} = K_1 \left( \sin^2 \Theta_H - \sin^2 \Theta_0 \right) \qquad (2)$$

where $\Theta_H$ - the angle between the *a*-axis and the magnetization vector in the field, $\Theta_0$ - the angle between the *a*-axis and the magnetization vector without the field.

Thus, the magnetic contribution to entropy caused by the process of rotation of the magnetization vector in the prismatic plane is:

$$\Delta S_{anis} = -\frac{\partial K_1}{\partial T} (\sin^2 \Theta_H - \sin^2 \Theta_0). \qquad (3)$$

As a rule, in the temperature region near Curie temperature, the anisotropy field is small and the magnetization vector completely rotates in the direction of the magnetic field, therefore, in our case, from equation (3) we can get an equation for RMCE:

$$\Delta S_{rot} = -\frac{\partial K_1}{\partial T} \qquad (4)$$

As can be seen from equation (4), the value of rotational entropy is determined by the temperature behavior of the first anisotropy constant. For this reason, the observed minimum cannot be explained within the framework of the classical theory of magnetocrystalline anisotropy. Earlier studies RMCE near the Curie temperature was carried out in [28]. One broad maximum of the RMCE was observed much lower than the Curie temperature. An increase in the magnetic field led to an expansion of the maximum of the RMCE. In this case, the existence of a complex temperature dependence of the RMCE can be associated with the anisotropy of the paramagnetic susceptibility. In work [3] at temperatures above the Curie temperature, it was shown that the magnetic susceptibility has a strong dependence on the orientation of the single crystal in a magnetic field. In work [29], the influence of the orientation of the monoclinic $Nd_2Ti_2O_7$ single crystal in the magnetic field on the value of the paramagnetic Curie point was found. This phenomenon itself could lead to the appearance of two maxima in the temperature dependence of the RMCE in our sample.

To use expression (4), it is necessary to know the value of the magnetocrystalline anisotropy constants, which can be obtained from the magnetization curves along the hard axis of magnetization. In this paper, by definition, we consider:

$$\Delta T_{rot} = \frac{-T}{C_P(T,H)} \Delta S_{rot} \quad (5)$$

Therefore, we can use the MCE data obtained from direct measurements (Fig. 2) and the $C_P(T)$ data measured at the same magnetic field configurations (Fig. 3). Figure 3 shows the $C_P(T)/T$ dependence at $H = 0$ and at 18 kOe for two orientations of the crystal with respect to the applied magnetic field. It can be seen that the application of a 18 kOe field almost completely suppresses the anomaly of $\Delta T_{rot}(T)$. The orientation of the magnetic field in the direction of $H \perp a$ caused a change in the temperature dependence and a shift of the maximum temperature by 4 K towards low temperatures in comparison with the curve at $H||a$, i.e. the slight anisotropy is observed. By taking into consideration of the data shown in Fig. 3, $\Delta S_{rot}(T)$ can be written in the following expression:

$$\Delta S_{rot}(T) = \Delta S_{H||a}(T) - \Delta S_{H \perp a}(T) = \frac{1}{T}\left(C_P^{H||a}(T,H) \cdot \Delta T_{H||a}(T) - C_P^{H \perp a}(T,H) \cdot \Delta T_{H \perp a}(T)\right) \quad (6)$$

The result of the estimated rotating magnetic entropy change ($\Delta S_{rot}$) as a function of temperature in a magnetic field of 18 kOe using expression (6) is presented in the inset of Fig. 3. The temperature trend of $\Delta S_{rot}$ (the inset of Fig. 3) is similar to that of $\Delta T_{rot}$ (Fig. 3(b)).

We have studied the specific heat and magnetocaloric properties of $Cu_2MnBO_5$ over the temperature range of 60–350 K and in magnetic fields up to 18 kOe. At temperatures below the $T_C$, $C_P(T)/T$ shows a linear temperature-dependent behavior, which is associated with the predominance of two-dimensional antiferromagnetic interactions of magnons with a linear dispersion relation propagating in the walls of the ludwigite. The temperature-independent behavior $C_P/T=f(T)$ observed in the temperature range of 95–160 K is likely associated with the excitation of the Wigner glass phase. Due to its strong magnetocrystalline anisotropy, the MCE anisotropy, and consequently the RMCE, is observed in $Cu_2MnBO_5$. The deep minimum in RMCE near the $T_C$ is also observed and suggested to be associated with the anisotropy of the paramagnetic susceptibility.

**All authors contributed equally this work**

xThe work was supported by the Russian Science Foundation under Grant No. 18-12-00415. The research was also carried out as part of the state task of the Ministry of Science of the Russian Federation (# AAAA - A17-117021310366-5).

The data that support the findings of this study are available from the corresponding author upon reasonable request.

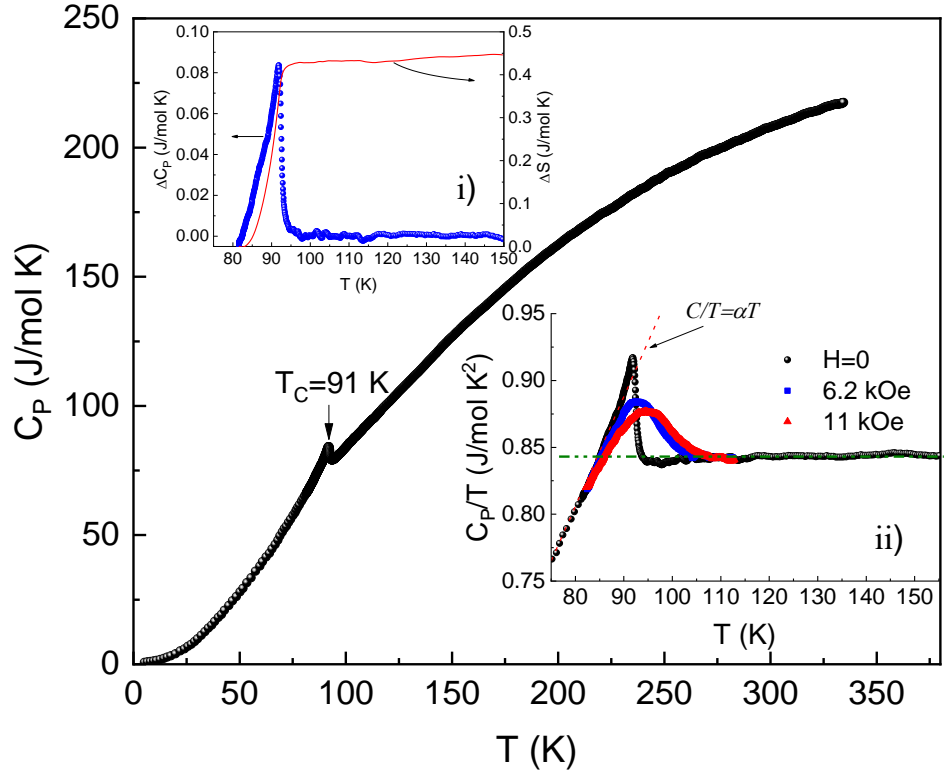

Fig. 1. Temperature dependence of $C_P/T$ for Cu$_2$MnBO$_5$ at $H$ = 0, 6.2, and 11 kOe. Inset (i) shows the temperature dependences of $\Delta C_P$ (blue) and the entropy change $\Delta S$ (red line). Inset (ii) shows the temperature dependence of an enlarged anomalous part of the $C_P/T$ (points).

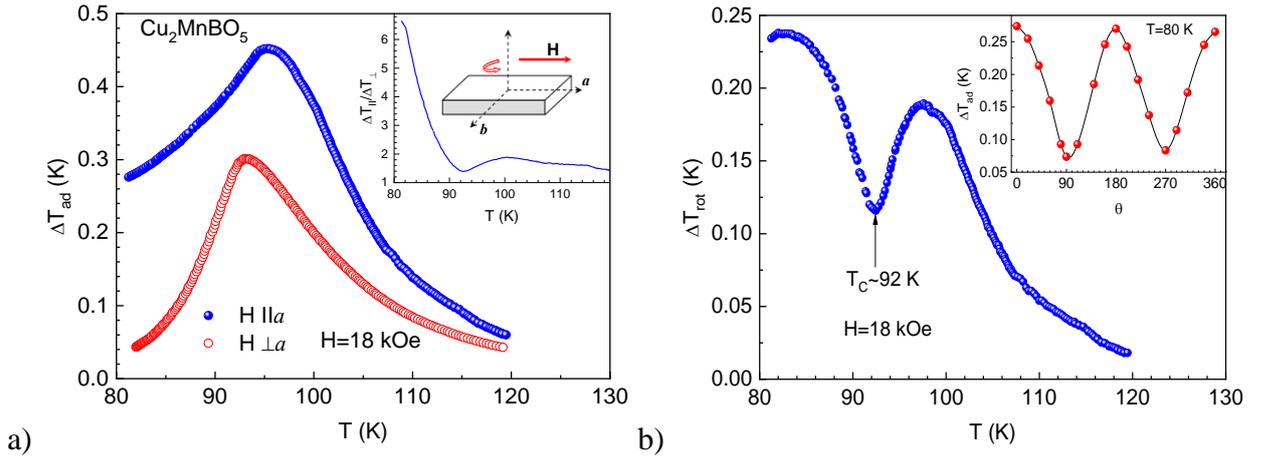

Fig. 2. a) Temperature dependence of MCE ($\Delta T_{ad}$) in a magnetic field of 18 kOe for two orientations of the crystal with respect to the applied magnetic field. The inset of Fig. 3(a) shows the $\Delta T_\parallel/\Delta T_\perp(T)$ dependence in a field of 18 kOe and the orientation of the crystal in a magnetic field. b) The adiabatic temperature change $\Delta T_{rot}$ of Cu$_2$MnBO$_5$ caused by a 90º rotation of the crystal as shown in the inset of Fig. 3(a). The inset of Fig. 3(b) shows the angular dependence of $\Delta T_{ad}$ at $T = 80$ K in a field of 18 kOe.

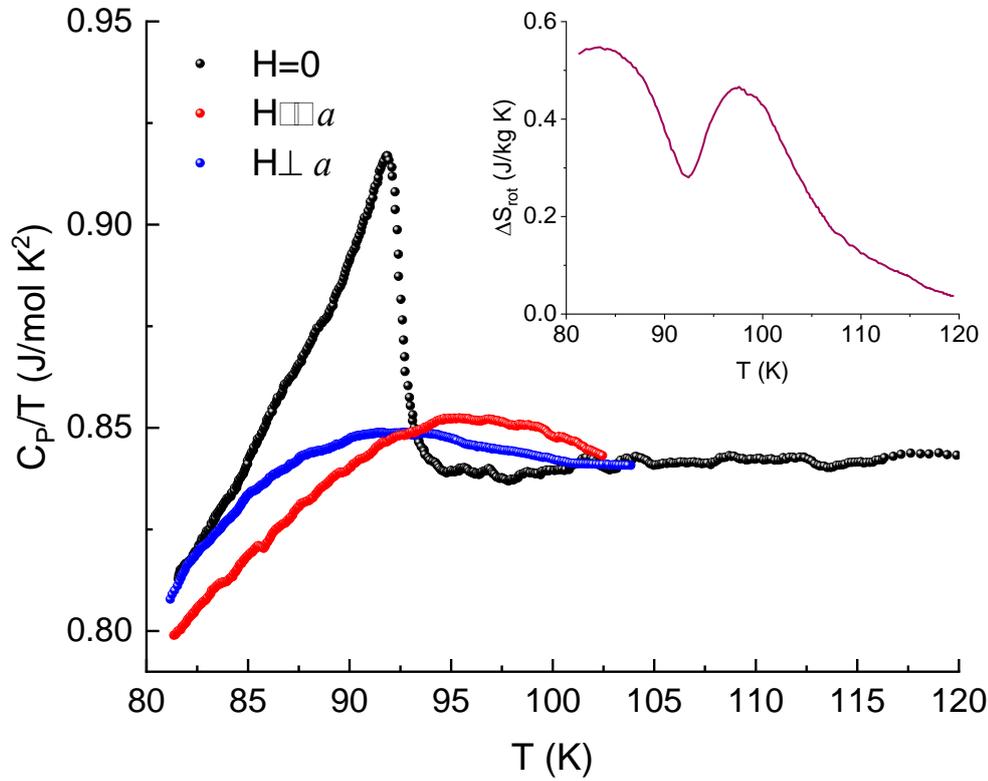

Fig. 3. $C_P(T)/T$ dependence at $H = 0$ and 18 kOe for two orientations of the crystal with respect to the magnetic field direction, $H||a$ and $H\perp a$. The inset shows the $\Delta S_{rot}(T)$ at $H = 18$ kOe.